\documentclass{aastex63}

\newcommand{\swift}{\textit{Swift}}
\newcommand{\xray}{\mbox{X-ray}}
\newcommand{\xrays}{\mbox{X-rays}}

\newcommand{\ergcms}{\mbox{erg cm$^{-2}$ s$^{-1}$}}
\def\deg{\hbox{$^\circ$}}
\def\arcmin{\hbox{$^\prime$}}
\def\arcsec{\hbox{$^{\prime\prime}$}}
\def\gtrsim{\ {\raise-.5ex\hbox{$\buildrel>\over\sim$}}\ }
\def\lesssim{\ {\raise-.5ex\hbox{$\buildrel<\over\sim$}}\ }

\received{November 2, 2020}
\submitjournal{ApJ}

\shorttitle{\swift\ follow-up of GW+neutrino}
\shortauthors{A. Keivani et al.}

\graphicspath{{./}{figures/}}

\begin{document}

\title{\swift\ \xray\ Follow-Up Observations of Gravitational Wave and High-Energy Neutrino Coincident Signals
}

\correspondingauthor{Azadeh Keivani}
\email{azadeh.keivani@columbia.edu}

\author[0000-0001-7197-2788]{Azadeh Keivani}
\affiliation{Department of Physics, Columbia University, New York, NY 10027, USA}
\affiliation{Columbia Astrophysics Laboratory, Columbia University, New York, NY 10027, USA}

\author[0000-0002-6745-4790]{Jamie A. Kennea}
\affiliation{Department of Astronomy \& Astrophysics, Pennsylvania State University, University Park, PA 16802, USA}

\author[0000-0002-8465-3353]{Phil A. Evans}
\affiliation{Department of Physics \& Astronomy, University of Leicester, Leicester, LEI 7RH, UK}

\author[0000-0002-2810-8764]{Aaron Tohuvavohu}
\affiliation{Department of Astronomy \& Astrophysics, Pennsylvania State University, University Park, PA 16802, USA}
\affiliation{Department of Astronomy \& Astrophysics, University of Toronto, Toronto, ON M5S 3H4 Canada}

\author{Riki Rapisura}
\affiliation{Department of Physics, Columbia University, New York, NY 10027, USA}

\author{Stefan Countryman}
\affiliation{Department of Physics, Columbia University, New York, NY 10027, USA}

\author[0000-0001-5607-3637]{Imre Bartos}
\affiliation{Department of Physics, University of Florida, Gainesville, FL 32611}

\author[0000-0003-1306-5260]{Zsuzsa M\'arka}
\affiliation{Columbia Astrophysics Laboratory, Columbia University, New York, NY 10027, USA}

\author[0000-0003-4225-0895]{Do\u{g}a Veske}
\affiliation{Department of Physics, Columbia University, New York, NY 10027, USA}

\author[0000-0002-3957-1324]{Szabolcs M\'arka}
\affiliation{Department of Physics, Columbia University, New York, NY 10027, USA}
\affiliation{Columbia Astrophysics Laboratory, Columbia University, New York, NY 10027, USA}

\author[0000-0002-3714-672X]{Derek~B. Fox}
\affiliation{Department of Astronomy \& Astrophysics, Pennsylvania State University,
  University Park, PA 16802, USA}
\affiliation{Center for Particle \& Gravitational Astrophysics, Institute for Gravitation and the Cosmos, Pennsylvania State University,
  University~Park, PA 16802, USA}
\affiliation{Center for Theoretical \& Observational Cosmology, Institute for Gravitation and the Cosmos, Pennsylvania State University, University Park, PA 16802, USA}


\begin{abstract}
Electromagnetic observations of gravitational wave and high-energy neutrino events are crucial in understanding the physics of their astrophysical sources.
\xray\ counterparts are especially useful in studying the physics of the jet, the energy of the outflow, and the particle acceleration mechanisms in the system.
We present the Neil Gehrels \swift\ Observatory prompt searches for \xray\ 
counterparts to the joint gravitational wave and high-energy neutrino coincident events that happened during the third observing run of LIGO/Virgo.
\swift\ observed the overlap between gravitational wave and neutrino error regions for three of the considerable (p-value $<$ 1\%) joint gravitational wave and high-energy neutrino coincident alerts, which were generated by the IceCube Neutrino Observatory in realtime after triggering by the LIGO/Virgo gravitational wave public alerts.
The searches did not associate any \xray\ 
counterparts to any of the joint gravitational wave and high-energy neutrino coincident events, however, the follow-up of these alerts significantly improved the tiling techniques covering regions between the gravitational wave sky maps and neutrino's error regions, making the realtime system ready for the future potential discoveries. 
We will discuss the details of each follow-up procedure, the results of each search, and the plans for future searches.
\end{abstract}


 \section{Introduction} 
 \label{sec:intro}
Multi-messenger searches have opened up new doors to astrophysical source discoveries, which result in understanding the physics of the sources and the underlying mechanisms that produce the messenger particles and waves.
Notably, the GW170817 event~\citep{2017PhRvL.119p1101A} which was detected by the LIGO and Virgo detectors~\citep{2015CQGra..32g4001L,2015CQGra..32b4001A} and was immediately found to be in coincidence with the GRB-170817A~(\citet{2017ApJ...848L..14G,2017ApJ...848L..15S}), started an extensive multi-messenger search resulting in the discovery of a kilonova and in expanding our knowledge about binary neutron star systems~\citep{2017ApJ...848L..12A}. 
Around the same time, the IceCube-170922A neutrino candidate detected by the IceCube Neutrino Observatory found to be associated with a flaring blazar, TXS~0506+056, which also initiated broad follow-up observations in different wavelengths~\citep{2018Sci...361.1378I}.

Although we have observed a couple of significant multi-messenger events so far, we haven't yet been able to identify any significant joint sources of gravitational waves (GWs), neutrinos, and electromagnetic waves \citep{2020arXiv200402910A,2017ApJ...850L..35A,2020ApJ...898L..10A,2019ApJ...870..134A,2017PhRvD..96b2005A,2016PhRvD..93l2010A,2014PhRvD..90j2002A,2011PhRvL.107y1101B}.
For the second observing period of LIGO/Virgo, the joint search for coincident gravitational wave and high-energy neutrino events was upgraded and performed in near-real time~\citep{2019arXiv190105486C}, and low-latency joint search continued for the LIGO/Virgo observing run O3 with two methods
~(\citet{2019ICRC...36..930K,2019ICRC...36..918H,2019PhRvD.100h3017B}). 
Triggered by LIGO/Virgo's public alerts, the realtime analyses are run in IceCube to search for spatially correlated neutrinos within $\pm 500$ s time window around the GW trigger time \citep{2011APh....35....1B}.
If a p-value $< 0.01$ is achieved from the IceCube analysis, a GCN circular encourages the community for further follow-up observations in the direction of the most significant correlated neutrino~\citep{2020arXiv200402910A}.

The Neil Gehrels \swift\ Observatory followed up three of the coincident GW+neutrino candidates during O3 under our \swift\ Cycle~15 guest investigator program. 
The goal of these observations were to identify electromagnetic counterparts to GW+neutrino events.
Here, we report the results of these searches as well as the techniques we have developed to define the most probable parts of the joint GW+neutrino candidate. An overview of pointed \swift\ followup of all GW triggers in O3 is given in \cite{2020MNRAS.tmp.2829P} and Oates et al., (in prep), for XRT and UVOT respectively.

In Sec.~\ref{sec:obs}, we will present details of each \swift\ follow-up observation after reviewing the general properties of the respective gravitational wave and neutrino events, for all candidates.
In Sec.~\ref{sec:tech}, the techniques we used to perform the tiled observations with \swift\ will be presented.
We will discuss the likelihood of these GW+neutrino alerts to be real coincident events based on different possible source scenarios capable of producing joint GW+neutrino alerts, in the last section (Sec.~\ref{sec:disc}). 

\section{Observations}
\label{sec:obs}
In this section, three different gravitational wave events (S190728q, S191216ap, S200213t) detected by LIGO/Virgo will be reviewed along with the respective neutrino candidates from IceCube and the follow-up observations performed by \swift. The \swift-XRT data were analysed using the standard
GW analysis pipeline (see \citealt{2016MNRAS.462.1591E,2019ApJS..245...15K}; Page et al., in prep. For details of the source detection methodology, see \citealt{2020ApJS..247...54E}).

\subsection{LIGO/Virgo S190728q Candidate}
The LIGO Scientific Collaboration and the Virgo Collaboration (LVC) reported the identification of the compact binary merger candidate S190728q on 2019 July 28. 
The estimated false alarm rate of the event calculated through the LVC online analysis was determined to be $2.5\times 10^{-23}$~Hz, or about one in $10^{15}$ years ~\citep{2019GCN.25187....1L}.
Initially, the gravitational wave signal was classified to have the following probabilities: MassGap\footnote{The masses of the lightest object ranging between 2.5 and 5 solar masses, which are the masses of the heaviest observed neutron star and the lightest black hole, respectively.} (52\%), Binary Black Hole (BBH; 34\%), Neutron Star - Black Hole (NSBH; 14\%), Binary Neutron Star (BNS; $<1\%$), or Terrestrial ($<1\%$). 
Assuming an astrophysical origin for the event, it was calculated for the lighter object to have mass of less than three solar masses.
Using the \texttt{bayestar.fits.gz} sky map, the 90\% credible region was estimated to be 543 deg$^2$.
However upon further analysis, the gravitational wave classification was modified to the following probabilities: BBH (95\%), MassGap (5\%), NSBH ($<1\%$), or BNS ($<1\%$). With the new preferred sky map \texttt{LALInference.offline.fits.gz}, the 90\% credible region was also updated to be 104 deg$^2$, and there is strong evidence that the lighter object may not be lighter than three solar masses \citep{2019GCN.25208....1L}.
The GW alert triggered several observatories in real-time and many follow-up observations were performed which included the search for low-energy neutrino candidates by the IceCube Neutrino Observatory.\\

\textbf{IceCube Neutrino Candidate:}
IceCube searched its data for track-like muon neutrino candidates within a 1000-second time span centered around the GW candidate S190728q alert time of 06:45:10.529 UTC on 2019 July 28 ~\citep{2019GCN.25185....1I}. 
In the initial search, no significant track-like events were captured by IceCube during this timeframe. 
However, further analysis based on the updated GW skymap found one neutrino candidate in spatial (RA, Dec = 312.87\deg, 5.85\deg; J2000) and temporal (time offset of $-360$~s) coincidence with the GW candidate S190728q~\citep{2019GCN.25210....1I}. 
The 90\%-containment radius for this neutrino candidate was 4.81\deg.
Overall, two hypothesis tests were conducted on the neutrino candidate. 
One utilized a maximum likelihood analysis which yielded an overall p-value of 0.014 ($2.21 \sigma$; \cite{2019ICRC...36..918H}), while the other utilized a Bayesian approach which yielded an overall p-value of 0.01 ($2.33 \sigma$; \cite{2019PhRvD.100h3017B}).
These p-values are measured against the known atmospheric backgrounds.\\

\begin{figure}
\begin{center}
 \includegraphics[scale=0.35]{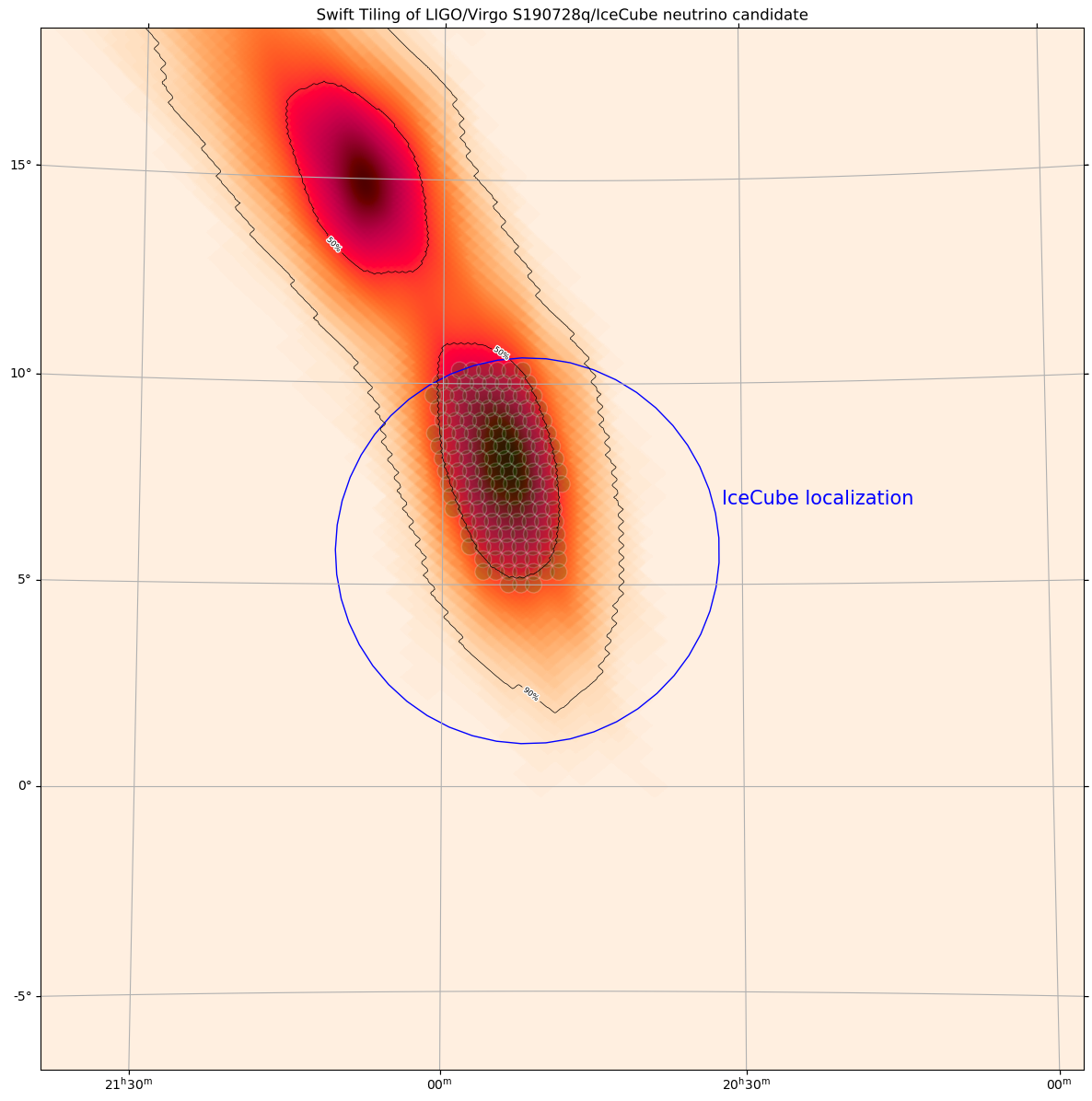}
\caption{Tiling map of \swift-XRT to follow up the joint LIGO/Virgo S190728q and IceCube neutrino candidate. The blue circle shows the 90\% containment circle of the IceCube neutrino candidate.} 
\label{fig:s190728q-tilingmap}
\end{center}
\end{figure}

\textbf{\swift\ Observations:}
\swift\ observed the direction of the IceCube track-like muon neutrino candidate~\citep{2019GCN.25210....1I}, which was consistent with the sky localization of gravitational-wave candidate S190728q~\citep{2019GCN.25208....1L}, covering $\sim14.5$ deg$^2$ in 145 tiles (Fig.~\ref{fig:s190728q-tilingmap}), to cover the most probable regions of the joint GW and neutrino localization~\citep{2019GCN.25220....1C}.
The observations ran from 2019 July  28 at 19:27 UT to 2019 July 29 at 11:41 UT, i.e.\ 46--104 ks after the LVC trigger. The average sensitivity of the observations was $\sim 2 \times 10^{-12}$ \ergcms\ (0.3$-$10 keV).

Three \xray\ sources were detected. The automated analysis pipelines assigns each source a `rank' of 1-4 which describes how likely it is to be related to the GW trigger, with 1 being the most likely and 4 being the least likely\footnote{https://www.swift.ac.uk/ranks.php}. All three sources matched known \xray\ sources.
We used a power-law spectrum with $N_H=3\times10^{20}$ cm$^{-2}$, and photon index $\Gamma=1.7$ to convert the measured count-rates to fluxes for comparison with catalogues; all three sources had fluxes below than their catalogued values, and were therefore assigned rank 4 and are not considered as likely counterparts to the joint GW+neutrino candidate. Details of these sources are listed in Table~\ref{tab:s190728q}.

\begin{deluxetable*}{cccccccc}
\tablecaption{\xray\ sources found in the \swift\ follow-up observations of the joint LIGO/Virgo S190728q and IceCube neutrino candidate. \label{tab:s190728q}}
\tablewidth{0pt}
\tablehead{
\colhead{Source} & \colhead{Match} & \colhead{RA} & \colhead{Dec} & \colhead{Err$_{90}$} & \colhead{$F_{12}$\tablenotemark{a}} & \colhead{Flag\tablenotemark{b}} & \colhead{Rank}}
\startdata
S190728q\_X1 & 1RXS J205242.6+081039   & 313.1761\deg\  &  8.1785\deg\ & 6.0\arcsec & 4.6 ($\pm$1.9) & Good & 4 \\
S190728q\_X2 & 1RXS J205421.7+090229   & 313.58716\deg\ & 9.0381\deg\  & 6.9\arcsec & 2.6 ($\pm$1.3) & Good & 4 \\
S190728q\_X3 & XMMSL2 J204928.9+060159 & 312.37177\deg\ & 6.0305\deg\  & 6.2\arcsec & 4.7 ($\pm$1.6) & Good & 4 \\
\enddata
\tablenotetext{a}{$F_{12}$ is 0.3-10 keV flux in units of $10^{-12} \ergcms$, calculated assuming a power-law spectrum with photon index of $\Gamma = 1.7$ and N$_{\rm{H}}=3\times 10^{20}$ cm$^{-2}$.}
\tablenotetext{b}{Flags Good/Reasonable/Poor refer to a spurious detection rate of $\sim$0.3\%, 1\% and 10\% respectively; see \cite{2020ApJS..247...54E} for details.}
\end{deluxetable*}

\subsection{LIGO/Virgo S191216ap Candidate}
On 2019 December 16, LVC reported the detection of the compact binary merger candidate S191216ap \citep{2019GCN.26454....1L}. 
The false alarm rate estimated by the online analysis was calculated to be about $1.1\times10^{-23}$ Hz, or one in $10^{15}$ years. 
The GW signal was initially classified with the following probabilities: MassGap ($>99\%$), BBH ($<1\%$), NSBH ($<1\%$), BNS ($<1\%$), or Terrestrial ($<1\%$). 
Assuming astrophysical origins, there was a 19\% probability that the lighter compact object had a mass that was less than three solar masses. 
The 90\% credible region was identified to be 300 deg$^2$ using the \texttt{bayestar.fits.gz,0} skymap. 
Further analysis was performed and the new classification of the event was described by the following probabilities: BBH ($>99\%$), MassGap ($<1\%$), Terrestrial ($<1\%$), BNS ($<1\%$), or NSBH ($<1\%$) \citep{2019GCN.26570....1L}. 
Along with this, the new preferred skymap was \texttt{LALInference.fits.gz,0} and the 90\% credible region was decreased to 253 deg$^2$ \citep{2019GCN.26505....1L}.
As with previous detections, the GW alert prompted other observatories to collect data in realtime. 
Follow-up searches of collected data were also performed by observatories such as the IceCube Neutrino Observatory.\\

\textbf{IceCube Neutrino Candidate:}
In a follow-up search by IceCube, one neutrino candidate was found in spatial (RA, Dec = 323.19\deg, 4.53\deg; J2000) and temporal (time offset of $-43$~s, with neutrino being detected before the GW alert) coincidence with the GW alert S191216ap \citep{2019GCN.26463....1C}. 
The IceCube search was initially planned to be performed in a 1000-second time window centered around the S191216ap alert time 21:34:01 UTC.
However, power issues occurred at the experimental site during the time of 21:33:21 UTC, which interfered with the data collection quality. 
Thus, only data collected before this time were considered. 
The neutrino candidate detected by IceCube underwent both the maximum likelihood and Bayesian analyses. 
These hypothesis tests obtained p-values of 0.104 (1.26$\sigma$) and 0.0059 (2.52$\sigma$), respectively.

It is also worth noting that 
the ANTARES Neutrino Observatory (with a lower sensitivity compared to IceCube for these candidates) did not detect any counterpart neutrino candidate for the S191216ap event \citep{2019GCN.26458....1A}.\\

\begin{figure}
\begin{center}
 \includegraphics[scale=0.25]{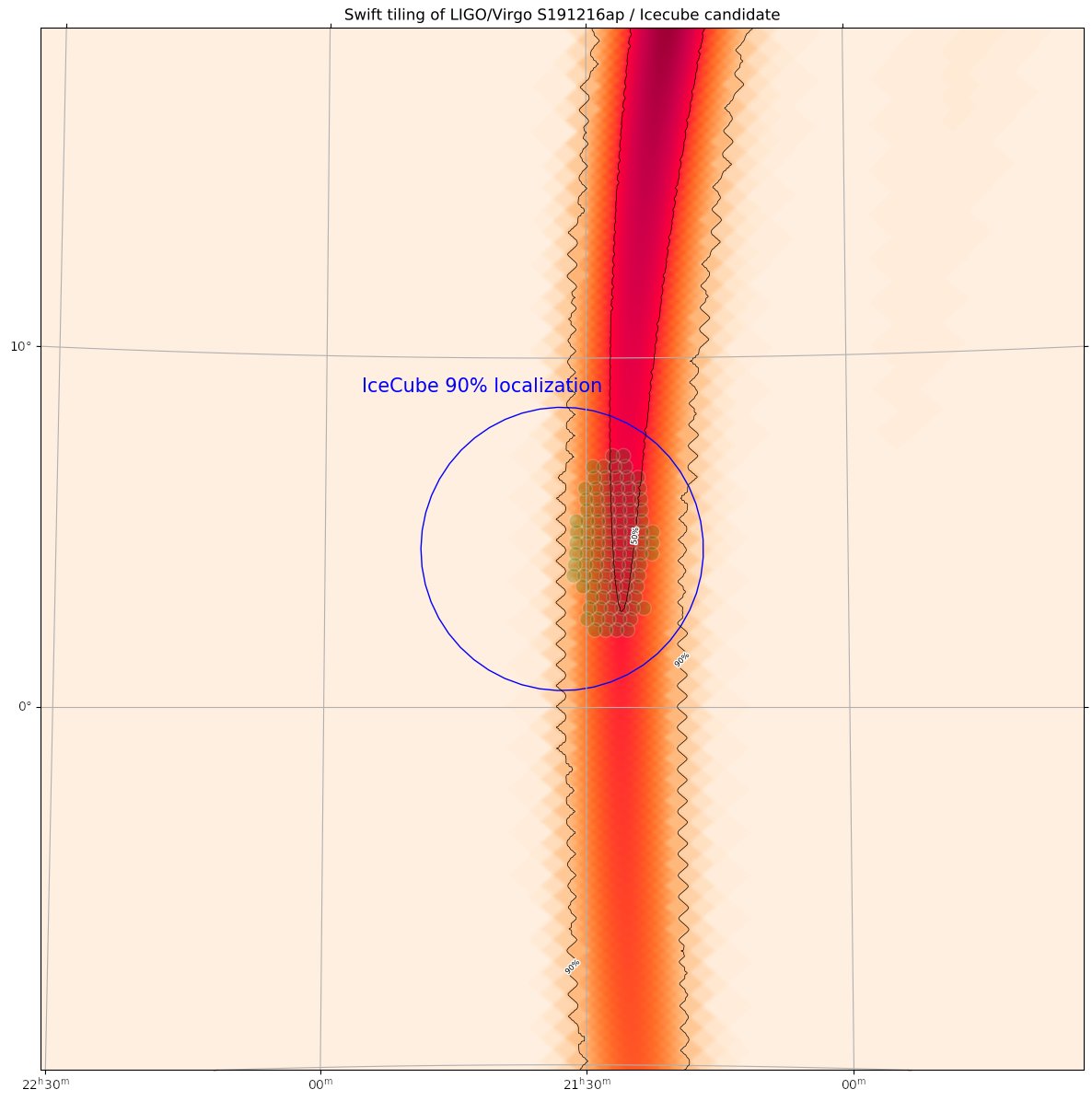}
\caption{Tiling map of \swift-XRT following up the joint LIGO/Virgo S191216ap and IceCube neutrino candidate. The blue circle shows the 90\% containment circle of the IceCube neutrino candidate.} 
\label{fig:s191216ap-tilingmap}
\end{center}
\end{figure}

\textbf{\swift\ Observations:}
\swift\ carried out 100 observations of the overlap between the LVC (\cite{2019GCN.26454....1L}) and the IceCube (\cite{2019GCN.26463....1C}) error regions for the GW trigger S191216ap~\citep{2019GCN.26475....1E}.
The observations were taken on 2019 December 17, from 03:56 UT to 09:14 UT; i.e.\ 23--42 ks after the LVC trigger. The average sensitivity of the observations was $\sim 2 \times 10^{-12}$ \ergcms\ (0.3$-$10 keV) (Fig.~\ref{fig:s191216ap-tilingmap}). 
This covered 3.3\% of the probability in the reported skymap using \texttt{bayestar.fits.gz}, 5.8\% after convolving with the 2MPZ galaxy catalogue, as described by \cite{2016MNRAS.462.1591E}, and 65\% of the probability contained within the combined GW and neutrino localizations. 
The pointings and associated metadata were also reported to the Treasure Map\footnote{http://treasuremap.space/alerts?graceids=S191216ap} (\citet{2020ApJ...894..127W,2019GCN.26244....1W}).

\swift-XRT detected three X-ray sources.  One of these was rank 3 (uncatalogued in \xrays, 
with flux below upper limit generated from existing surveys), and two of rank 4 (catalogued in X-rays, with a catalogued flux consistent with or brighter than our measured flux). 


Further searches were performed by \swift~\citep{2019GCN.26498....1E} after the High-Altitude Water Cherenkov (HAWC) Observatory revealed a relatively significant sub-threshold event at RA, Dec = 323.53\deg, 5.23\deg\ (J2000) coincident with the updated localization of the neutrino event~\citep{2019GCN.26472....1H}.
\swift\ observations covered the HAWC error region for $\sim$500~s per tile from 2019 December 19 at 18:18 UT to 2019 December 20 at 02:20 UT (i.e. 75--104 ks after the LVC trigger), with a sensitivity of $\sim 4\times10^{-13}$\ \ergcms.
Additionally, $\sim$3~ks observations were conducted for each of the galaxies listed by \cite{2019GCN.26479....1S}. These observations ran between 2019 December 19 at 23:41 UT and 2019 December 20 at 12:09 UT (i.e. 94--139 ks after the LVC trigger), with a sensitivity of $10^{-13}$\ \ergcms.
A total of 18 \xray\ sources were detected in these searches; 14 of rank 3 and four of rank 4; given these ranks, none of these are strong candidates for the counterpart to the GW event. Details of these sources are listed in Table~\ref{tab:s191216ap}.

\begin{deluxetable*}{cccccccc}
\tablecaption{\xray\ sources found in the \swift\ follow-up observations of the joint LIGO/Virgo S191216ap and IceCube neutrino candidate. \label{tab:s191216ap}}
\tablewidth{0pt}
\tablehead{
\colhead{Source} & \colhead{Match} & \colhead{RA} & \colhead{Dec} & \colhead{Err$_{90}$} & \colhead{$F_{12}$} & \colhead{Flag} & \colhead{Rank}}
\startdata
S191216ap\_X1 &                          & 322.1300\deg  & +4.8310\deg  & 6.1\arcsec  & 3.2$^{+2.3}_{-1.6}$        & Good & 3\\
S191216ap\_X2 & XMMSL2 J212817.4+034848  & 322.0752\deg  & +3.8154\deg  & 7.7\arcsec  & 6.7$^{+3.2}_{-2.4}$        & Good & 4\\
S191216ap\_X3 & 2SXPS J213053.0+040229   & 322.7186\deg  & +4.0414\deg  & 5.5\arcsec  & 6.5$^{+3.4}_{-2.5}$        & Good & 4\\
S191216ap\_X4 &                          & 323.4550\deg  & +4.9553\deg  & 6.7\arcsec  & 0.87$^{+0.47}_{-0.34}$     & Good & 3\\
S191216ap\_X5 &                          & 323.3284\deg  & +5.2847\deg  & 6.9\arcsec  & 0.55$^{+0.33}_{-0.23}$     & Good & 3\\
S191216ap\_X6 &                          & 323.7216\deg  & +5.3197\deg  & 6.6\arcsec  & 0.72$^{+0.38}_{-0.28}$     & Good & 3\\
S191216ap\_X7 &                          & 323.1842\deg  & +4.3350\deg  & 5.8\arcsec  & 0.13$^{+0.05}_{-0.04}$     & Good & 3\\
S191216ap\_X8 &                          & 323.1995\deg  & +4.4645\deg  & 5.5\arcsec  & 0.24$^{+0.08}_{-0.06}$     & Good & 3\\
S191216ap\_X9 &                          & 323.1114\deg  & +4.3588\deg  & 5.3\arcsec  & 0.28$^{+0.10}_{-0.08}$     & Good & 3\\
S191216ap\_X10 &                         & 323.0271\deg  & +4.3106\deg  & 6.9\arcsec  & 0.14$^{+0.07}_{-0.05}$     & Good & 3\\
S191216ap\_X11 & 1RXS J213242.1+042357   & 323.1709\deg  & +4.4052\deg  & 6.2\arcsec  & 0.16$^{+0.06}_{-0.05}$     & Good & 4\\
S191216ap\_X12 &                         & 322.9441\deg  & +5.7457\deg  & 7.0\arcsec  & 0.16$^{+0.07}_{-0.05}$     & Good & 3\\
S191216ap\_X13 &                         & 322.9867\deg  & +5.1896\deg  & 8.0\arcsec  & 0.10$^{+0.06}_{-0.05}$     & Good & 3\\
S191216ap\_X14 & 2SXPS J213032.7+050216  & 322.6371\deg  & +5.0384\deg  & 7.6\arcsec  & 1.1$^{+1.6}_{-0.8}$        & Good & 4\\
S191216ap\_X15 & 2SXPS J213026.0+050604  & 322.6096\deg  & +5.1020\deg  & 7.2\arcsec  & 0.16$^{+0.07}_{-0.06}$     & Good & 4\\
S191216ap\_X16 & 2SXPS J213122.3+050236  & 322.8443\deg  & +5.0437\deg  & 5.4\arcsec  & 0.20$^{+0.05}_{-0.05}$     & Good & 4\\
S191216ap\_X17 &                         & 322.9324\deg  & +5.4210\deg  & 6.0\arcsec  & 0.17$^{+0.08}_{-0.06}$     & Good & 3\\
S191216ap\_X18 &                         & 322.7879\deg  & +5.5795\deg  & 4.6\arcsec  & 0.21$^{+0.08}_{-0.06}$     & Good & 3\\
S191216ap\_X19 &                         & 322.9901\deg  & +4.7273\deg  & 7.4\arcsec  & 0.13$^{+0.08}_{-0.06}$     & Good & 3\\
S191216ap\_X20 &                         & 322.8833\deg  & +4.9008\deg  & 5.0\arcsec  & 0.31$^{+0.10}_{-0.08}$     & Good & 3\\
S191216ap\_X21 &                         & 322.9354\deg  & +5.3867\deg  & 7.4\arcsec  & 1.1$^{+1.7}_{-0.8}$        & Poor & 3\\
\enddata
\tablecomments{See Table~1 for column definitions.}
\end{deluxetable*}

\subsection{LIGO/Virgo S200213t Candidate}
The LVC observed the compact binary merger candidate S200213t on 2020 February 13 \citep{2020GCN.27042....1L}. 
The detected GW event had a false alarm rate of approximately $1.8\times10^{-8}$ Hz, or once every one year and nine months. 
The classification of the GW signal was described by the following probabilities: BNS (63\%), Terrestrial (37\%), BBH ($<1\%$), MassGap ($<1\%$), or NSBH ($<1\%$). 
The probability that the lighter compact object is lighter than three solar masses was $>99$\% when assuming astrophysical origins.
Additionally, the 90\% credible region was identified as 2587 deg$^2$ when using the preferred \texttt{bayestar.fits.gz,1} skymap. 
Another analysis was performed and the preferred skymap was updated to be the \texttt{bilby.fits.gz,0} skymap, while the 90\% credible region was 282 deg$^2$ \citep{2020GCN.27092....1L}. However, another round of analysis revealed a new preferred sky map, \texttt{LALInference.fits.gz,0} and a new 90\% credible region of 2326 deg$^2$ \citep{2020GCN.27096....1L}.
LVC reported that the last Circular superseded the previous Bilby analysis, which was based on an outdated estimate of the detector calibration uncertainty.
Upon detection, the GW alert initiated other observatories such as the IceCube Neutrino Observatory and the Pierre Auger Observatory to look for neutrino detections using realtime analyses.\\

\textbf{IceCube Neutrino Candidate:}
IceCube detected one neutrino candidate that was in coincidence with GW alert S200213t both spatially (RA, Dec = 45.21\deg, 31.74 \deg; J2000) and temporally (-175.94~s). 
The neutrino candidate underwent both the maximum likelihood and Bayesian analyses. 
These hypothesis tests obtained p-values of 0.003 ($2.75 \sigma$) and 0.0174 ($2.11 \sigma$), respectively~\citep{2020GCN.27043....1I}.
The 90\%-containment radius for this neutrino candidate was estimated as 0.43\deg.

No neutrino candidates were detected by the ANTARES detector and the Pierre Auger Observatory during a $\pm$500~s interval from the S200213t alert time of 04:11:04 UTC on 2020 February 13 (\citet{2020GCN.27049....1A,2020GCN.27076....1A}). 

Utilizing the obtained localization area of the joint GW+neutrino candidate detection area, follow-up observations for host galaxy candidates were conducted~\citep{2020GCN.27119....1P}. 
The Lemonsan Optical Astronomical Observatory, the Deokheung Optical Astronomy Observatory~(DOAO), and the Sobaeksan Optical Astronomy Observatory (SOAO) examined these host galaxies but found no electromagnetic counterparts~\citep{2020GCN.27119....1P}.\\

\begin{figure}[th]
\begin{center}
 \includegraphics[scale=0.4]{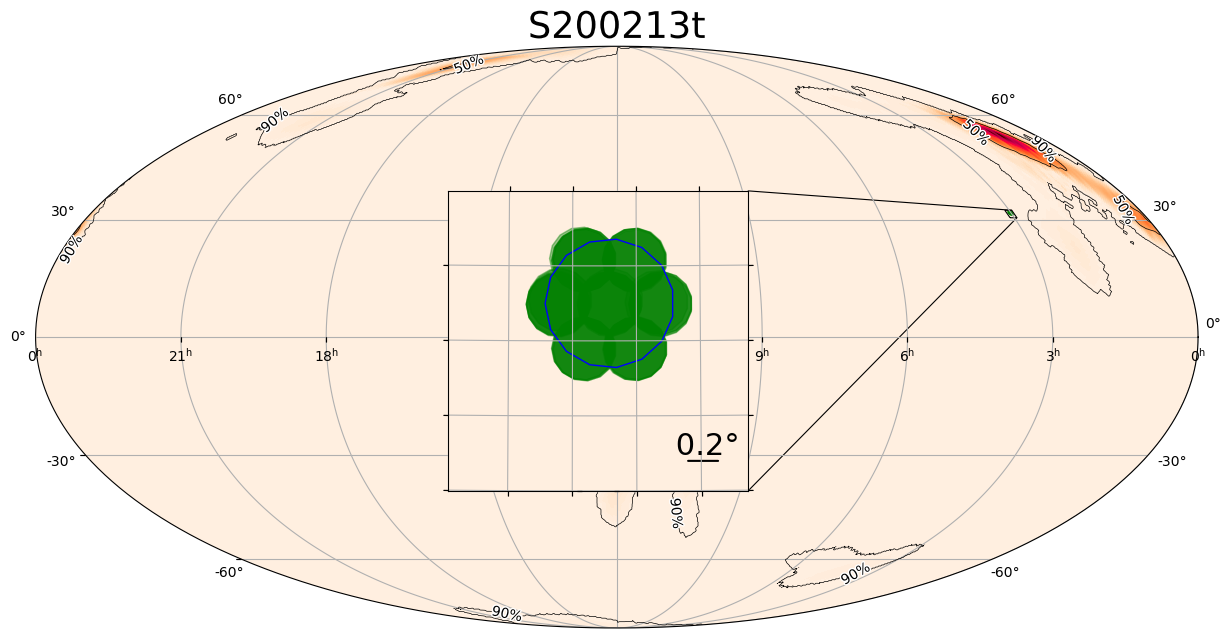}
\caption{Tiling map of \swift-XRT to follow up the joint LIGO/Virgo S200213t and IceCube neutrino candidate. The green circles show the \swift\ 7-point tiling pattern, centered around the location of the IceCube neutrino candidate.} 
\label{fig:s200213t-tilingmap}
\end{center}
\end{figure}

\textbf{\swift\ Observations:}
\swift\ observed the field of the IceCube track-like muon neutrino candidate~\citep{2020GCN.27043....1I} consistent with the sky localization of GW candidate S200213t~\citep{2020GCN.27042....1L}, covering $\sim 0.5$ deg$^2$ in seven tiles to cover the most probable regions of the joint GW+neutrino localization~\citep{2020GCN.27121....1C}.
The observations were taken on 2020 February 13, from 09:54 UT to 16:20 UT; i.e.\ 21--44 ks after the LVC trigger. The average sensitivity of the observations was $\sim 8 \times 10^{-13}$ \ergcms\ (0.3$-$10 keV) (Fig.~\ref{fig:s200213t-tilingmap}).  
Since the 90\%-containment radius for this neutrino candidate was 0.43\deg, there was no need to find the convolved skymaps of GW and neutrino events; instead \swift\ used one of its on-board tiling patterns to observe the location of the IceCube neutrino candidate coincident with the GW skymap.
The typical sensitivity for this search was $2\times 10^{-13}$ \ergcms (0.3-10 keV).

In total three X-ray sources were found with detection flag of ``good", all of which were classified as ``rank 3", i.e. uncatalogued in X-rays, but with fluxes below historical upper limits. 
Two further ``rank 3" sources were found with detection flag of ``poor", indicating that they are likely spurious.
Details of the ``good" sources are listed in Table~\ref{tab:S200213t}.\\

\begin{deluxetable*}{ccccccc}
\tablenum{3}
\tablecaption{\xray\ sources found in the \swift\ follow-up observations of the LIGO/Virgo S200213t and IceCube neutrino candidate. \label{tab:S200213t}}
\tablewidth{0pt}
\tablehead{
\colhead{Source} &  \colhead{RA} & \colhead{Dec} & \colhead{Err$_{90}$} & \colhead{$F_{12}$} & \colhead{Flag} & \colhead{Rank} }
\startdata
S200213t\_X2 & 45.3486\deg & +31.7006\deg & 5.8\arcsec & $0.8\pm0.2$            & Good & 3\\
S200213t\_X3 & 45.1090\deg & +31.9198\deg & 6.6\arcsec & $0.^{+0.10}_{-0.08}$   & Poor & 3\\ 
S200213t\_X4 & 44.9937\deg & +31.4714\deg & 7.1\arcsec & $0.22^{+0.14}_{-0.10}$ & Good & 3 \\
S200213t\_X5 & 45.1234\deg & +31.9529\deg & 5.3\arcsec & $0.29^{+0.12}_{-0.09}$ & Poor & 3 \\
S200213t\_X6 & 45.0670\deg & +31.9802\deg & 4.7\arcsec & $0.23^{+0.15}_{-0.10}$ & Good & 3\\
\enddata
\tablecomments{See Table~1 for column definitions.}
\end{deluxetable*}

\section{Techniques}
\label{sec:tech}
One of the great advantages of the GW+neutrino searches comes from the fact that neutrinos have far smaller angular uncertainties compared to the GW events, which provides a great opportunity for observatories with smaller fields of view to search for electromagnetic counterparts of gravitational waves.

The ideal case to follow-up a joint GW+neutrino candidate event with \swift\ happens when the neutrino is relatively well-localized, i.e. its 90\%-containment radius is \lesssim{1\deg}.
In such cases, \swift\ can easily use one of its on-board tiling patterns to conduct follow-up searches: 7 tiles for 12\arcmin\ $<$ R $<$ 33\arcmin, 19 tiles for 33\arcmin\ $<$ R $<$ 47\arcmin, or 37 tiles for 47\arcmin\ $<$ R $<$ 66\arcmin, where R is the neutrino's 90\%-containment radius. 
In cases that the neutrino's 90\%-containment radius is {$>$ 1\deg}, the \swift+LVC+IceCube team needs to make a decision on how many tiles to dedicate for the follow-up of the alert based on the joint GW+neutrino probability density map and the observing priorities.


In the case where error regions are significantly larger, or not optimally covered by the nominally hexagonal \swift\ tiling techniques, \swift\ has the ability to perform large-scale arbitrary tiling, which was first developed for in order to meet the needs of covering the large tiling areas required for follow-up GW-only triggers \citep{2016MNRAS.462.1591E,2018IAUS..338...53T}. In the case of a LVC+IceCube trigger, the convolved probability map is taken, and an optimal tiling solution is calculated by placing overlapping tiles on the probability map. This starts by defining a grid of fix spaced tiles, centered on the region of highest probability. Tiles are added at these fixed positions which cover the regions of highest probability. The fixed spacing of the tiles means that there is overlap between them, which ensures full coverage of the probability region without gaps which might otherwise occur when covering a uniform region with the approximately circular \swift/XRT field of view. For small circular error regions, this algorithm reproduces the 7-, 19- and 37-tile automated tiling regions, with the added benefit that it can handle much larger and non-uniformly shaped error regions. The tiling algorithm continues to add tiles until either a defined maximum number of tiles is reached, or until the total integrated probability covered by the tiles reaches a maximum value. 

The list of tile coordinates is then fed into a high fidelity observation planning algorithm, which based upon the current \swift\ observing plan and the visibility of the tiles, calculates an optimized observing timeline, prioritizing both speed of completion and coverage of the highest probability regions earliest. Due to the complexity of these tiling plans, they require a ground station contact to upload to \swift\ before execution can begin. Typical latencies for \swift\ ground station contacts is around 40 minutes, although longer gaps can occur. In the case of S190728q and S191216ap this tiling method was utilized, and the tiling patterns that resulted can be seen in Figures~\ref{fig:s190728q-tilingmap} and \ref{fig:s191216ap-tilingmap}.


\section{Discussion}
\label{sec:disc}
Table~\ref{tab:summary} summarizes the information about the three GW+neutrino candidates that were followed up by \swift\ and were discussed in this paper.
While it turned out that the first two GW candidates (S190728q and S191216ap) in this study were most likely from BBH mergers, the last one, S200213t, was categorized as a BNS with 63\% probability.

In a BNS merger event, it is thought that high energy neutrinos are produced in non-thermal dissipative processes, such as relativistic outflows~\citep{KohraBartos,2013CQGra..30l3001B} and could produce short gamma-ray bursts (GRB) within two seconds of the merger. 
The case of GW170817/GRB170817 was a strong evidence for the presence of such systems. 
No neutrinos were found to be spatially coincident with the GW170817 event \citep{2017ApJ...850L..35A}, within $\pm 500$\,s of the merger.
This was consistent with a short GRB observed at a large off-axis angle~\citep{2017ApJ...850L..35A}.
For the case of S200213t, there is no short GRB associated with the GW event. 

Although it is more likely for neutrinos to be produced in a compact binary system including one neutron star (i.e. a BNS or a NSBH), it is still possible to detect neutrinos from a BBH merger if the black holes are in a gas rich environment where they can accrete from \citep{KohraBartos}. 
Such a scenario could arise within the accretion disk of active galactic nuclei \citep{2017ApJ...835..165B,2017MNRAS.464..946S,2019ApJ...884L..50M,2019ApJ...876..122Y,2019PhRvL.123r1101Y,2020ApJ...896..138Y,2020arXiv200704781Y}. A candidate counterpart for a black hole merger has already been detected by the Zwicky Transient Facility, which, if true, would support this emission scenario \citep{2020PhRvL.124y1102G}. Nonetheless, these systems need further studies to better understand their properties and the expected electromagnetic and neutrino emission from them.

The highest statistical significance among our alerts was $2.75\sigma$ for the S200213t GW event, which is not high enough to claim a significant coincidence.
Moreover, the neutrino was detected 176 seconds before the GW event, which is probably before the BNS started to distort. 
This time offset reduces the likelihood of the joint signal to be from the same source.

While this search did not conclusively identify a multi-messenger counterpart, the study demonstrated the feasibility and need for such follow-ups. 
Improving statistical methods could be extremely useful in searching for multiple messengers ~\citep{2020arXiv201004162V}.
For LIGO/Virgo/KAGRA's upcoming O4 observing run, the number of expected detections will dramatically increase compared to O3. In addition, the discovery of the black hole merger GW190521 \citep{GW190521discovery} and its candidate electromagnetic counterpart \citep{2020PhRvL.124y1102G} could further substantially increase the interest in the multi-messenger follow-ups of binary black hole mergers. 
Other than the counterpart, the properties of GW190521 also point towards an AGN origin \citep{GW190521_properties}, including its possibly high eccentricity \citep{2020arXiv200905461G}. Therefore, the extension of the \swift\ GW+neutrino follow-up mission, along with the optical and high-energy follow-ups of possible \swift+GW+neutrino candidates, will be an exciting science target for the O4 observing run.\\

\begin{deluxetable*}{ccccccccc}
\tablenum{4}
\tablecaption{Summary of the GW+neutrino ($\nu$)~candidates \label{tab:summary}}
\tablewidth{0pt}
\tablehead{
\colhead{GW} & \colhead{GW alert time} & \colhead{Classification} & \colhead{GW$\_CR90$} & \colhead{RA$_\nu$} & \colhead{Dec$_\nu$} & \colhead{err90$_\nu$} & \colhead{dt} & \colhead{significance}}
\startdata
S190728q & 2019 July 28, 06:45:10.529 UTC & BBH & 104 deg$^2$ & 312.87\deg & 5.85\deg & 4.81\deg & $-360$~s & 2.33$\sigma$\\
S191216ap & 2019 Dec 16, 21:34:01 UTC & BBH & 253 deg$^2$ & 323.19\deg & 4.53\deg & 4.07\deg & $-43$~s & 2.52$\sigma$\\
S200213t & 2020 Feb 13, 04:11:04 UTC & BNS & 2326 deg$^2$ & 45.21\deg & 31.47\deg & 0.43\deg & $-176$~s & 2.75$\sigma$\\
\enddata
\tablecomments{The subscript $\nu$ denotes information about the neutrino candidate. GW\_CR90: 90\% credible region of the GW event; err90$_\nu$: 90\%-containment radius of the neutrino event; dt: time offset of track event with respect to the GW trigger. In the last column, the highest significance from the two searches (the Bayesian search and the maximum likelihood analysis) is reported here.}
\end{deluxetable*}

\textbf{Acknowledgements.} The authors thank the LVC and IceCube collaboration for publicly distributing their realtime alerts and the \swift\ team for the rapid response to our target of opportunity requests. 
A.K. gratefully acknowledges support from the National Aeronautics and Space Administration Swift Guest Investigator Program under grant 80NSSC20K0471.
AK acknowledges support from the Frontiers of Science fellowship at Columbia University.
The Columbia authors thank Columbia University in the City of New York for its unwavering support.
IceCube and LIGO were made possible with the generous support of the US National Science Foundation (NSF). 
I.B. acknowledges the support of the National Science Foundation under grant \#1911796 and of the Alfred P. Sloan Foundation. 
R.R. acknowledges the support from the Summer Enhancement Fellowship of Columbia Undergraduate Scholars Program.
D.V. acknowledges support from Fulbright foreign student program and Jacob Shaham Fellowship.

\bibliography{Swift_GW_Nu}{}
\bibliographystyle{aasjournal}

\end{document}